\documentclass[a4paper,11pt]{article}

\usepackage{jheppub} 
\usepackage[T1]{fontenc} 

\title{\boldmath Hidden Conformal Symmetry for Vector Field on Various Black Hole Backgrounds}

\author[a,c]{Changfu Shi}
\author[a,b]{Jian-dong Zhang}
\author[a,b]{Jianwei Mei}

\affiliation[a]{TianQin Research Center for Gravitational Physics, Sun Yat-sen University(Zhuhai Campus), Zhuhai 519082, China}
\affiliation[b]{School of Physics and Astronomy, Sun Yat-sen University(Zhuhai Campus), Zhuhai 519082, China}
\affiliation[c]{MOE Key Laboratory of Fundamental Physical Quantities Measurement, School of Physics, Huazhong
University of Science and Technology, 1037 Luo Yu Road, Wuhan 430074, China}

\emailAdd{cfshi@hust.edu.cn}
\emailAdd{zhangjd9@mail.sysu.edu.cn}
\emailAdd{meijw@sysu.edu.cn}

\abstract{Hidden conformal symmetries of scalar field on various black hole backgrounds have been investigated for years,	but whether those features hold for other fields are still open questions. Recently, with proper assumptions, Lunin achieved to the separation of variables for Maxwell equations on Kerr background.	In this paper, with that equation, we find that hidden conformal symmetry appears at near region under low frequency limit. We also extended those results to vector field on the more general Kerr-NUT-(A)dS background, then hidden conformal symmetry also appears if we focusing on the near-horizon region at low frequency limit.}

\begin{document}
\maketitle
\flushbottom

\section{Introduction}
\label{sec:intro}

One of the most mysterious features of black holes is the Bekenstein-Hawking Entropy.
To explain the area law of black hole entropy and to figure out its microscopic origin
are very important challenges for a quantum gravity theory.
A temptative explanation comes from the holographic principle proposed in \cite{tHooft:1993dmi} and \cite{Susskind:1994vu} which states that gravity can be described equivalently by a theory with one lower number of dimensions.
Then, the exact AdS/CFT correspondences \cite{Maldacena:1997re,Witten:1998qj} provided explicit realizations of the holographic principle.
This is one of the greatest achievements of modern theoretical physics.
Using holographic principle, the microscopic origin of the Bekenstein-Hawking entropy
was first reproduced by methods depend heavily on details of string theory \cite{Strominger:1996sh}.
Soon, it was understood in \cite{Strominger:1997eq} that it can be applied to any consistent, unitary quantum theory of gravity
which contains the black holes as classical solutions.

Although the AdS/CFT correspondence is very useful for studying those properties of black holes,
most of the black holes considered in the context of AdS/CFT are physically unrealistic.
Fortunately, about a decade ago, it was showed that the ideas of AdS/CFT could nevertheless be used to understand astrophysical Kerr black holes.
This is called as Kerr/CFT correspondence \cite{Compere:2012jk},
and it claims the duality between a Kerr black hole with mass \(M\) and angular momentum \(J=a M\)
and a 2D conformal field theory(CFT) with central charges \(c_L=c_R=12J\)
and temperatures \(T_L=\frac{M^2}{2\pi J},T_R=\frac{\sqrt{M^4-J^2}}{2\pi J}\).
This duality was first exhibited by \cite{Guica:2008mu} for extremal Kerr black holes with $J=M^2$.
An extremal Kerr black hole has a near-horizon scaling region which is known as the NHEK (Near-Horizon Extreme Kerr) geometry,
whose isometry group is \(SL(2, \mathbb{R})\times Á U(1)\).
Then, by choosing proper NHEK boundary conditions, the canonically conserved charges associated with the non-trivial diffeomorphisms
of the NHEK region form two copies of the two-dimensional Virasoro algebra with \(c_L = c_R = 12J\) \cite{Castro:2009jf,Bredberg:2009pv}.
One of the most important supports is that the Cardy microstate degeneracy on the CFT side precisely matched the Bekenstein Hawking entropy.
And such correspondence was soon generalized to various extremal or near-extremal black holes \cite{Lu:2008jk,Matsuo:2009sj,Hartman:2009nz,Cvetic:2009jn,Chen:2010ni,Carlip:2011ax,Lu:2009gj,Mei:2010wm,Mei:2012wd,Azeyanagi:2008kb,Hartman:2008pb,Chow:2008dp,Shi:2016jtn,Peng:2009ty,Compere:2009dp}.

Furthermore, the Kerr/CFT correspondence was extended to generic non-extremal case.
In a remarkable paper \cite{Castro:2010fd},
it is exhibited the existence of conformal symmetry in the solution space of the scalar wave equation.
This is called hidden conformal symmetry, and it allows us to associate a CFT description to general non-extremal black holes.
With the assumptions that the central charge will have the same form as extremal case,
the Cardy formula will reproduce the Bekenstein-Hawking entropy again.
As expressed in \cite{Chen:2010ik,Chen:2011dc}, it's an intrinsic property of the black hole instead of an artifact of the scalar equation of motion.
Further discussions about hidden conformal symmetry extended to various black holes can be found \cite{Chen:2010as,Wang:2010qv,Chen:2011kt,Chen:2010jj,Chen:2010bh,Chen:2010fr,Chen:2010bd,Bak:2011yy}.

But an acute problem is that all those analyzations are just for scalar field,
whether or not this hidden conformal symmetry could be found for higher spin waves are still open questions \cite{Compere:2012jk}.
Chen et al. \cite{,Chen:2010ik} analyzed the hidden conformal symmetry acts on vector and tensor field by defining a Lie-induced derivate. Lowe et al. \cite{Lowe:2013uea} tried to achieve this goal by analyzing the Teukolsky equation.
These results gave affirmative answers to this question, but none of them solved the problem in essence.

Recently, Lunin has done an excellent work on separation of variables for Maxwell equation in the Myers-Perry geometry \cite{Lunin:2017drx}.
He actually provided a useful frame to separate variables for higher spin fields  especially for Maxwell equations. We should notice that this is different from Teukolsky equation, since it's an equation of field instead of field strength.

In this paper, we analyzed the separation variables for Maxwell equation in Kerr and Kerr-NUT-(A)dS \cite{Chen:2006xh}  backgrounds with the ansatz provided by Lunin \cite{Lunin:2017drx}. Then we found the hidden conformal symmetries of Maxwell equation in those geometries.  In Kerr spacetime, this hidden conformal symmetry do exist in the near region under low frequency approximation, the separation equation can be reproduced by \(SL(2,\mathbb{R})\) Casimir just like the situation of scalar field. In  Kerr-NUT-(A)dS spacetime, we find that there also exist hidden conformal symmetry on the solution space of the Maxwell separation equations in the low-frequency limit if we focus on the near-horizon region. Extending those results to Kerr-Newman, Kerr-(A)dS, Kerr-Newman-(A)dS and Kerr-Newman-NUT-(A)dS are quite directly, and we have checked all those four situations. Since the logic are very similar with above discussions, we will not list them in this paper.

The paper is organized as follows.
In Sec.2, we will focus on Kerr black hole, first of all we will review the discussion of the hidden conformal  symmetry of scalar field briefly, and then discuss the hidden conformal symmetry of Maxwell equation on Kerr black hole background in the same way. In Sec.3, we will extend those results to Kerr-NUT-(A)dS geometry. In Sec.4, we will give a brief conclusion.

\section{Hidden conformal symmetry of Kerr black hole}

In this section, we will focus on the hidden conformal symmetry of Kerr black hole. First, we will review the discussion of hidden conformal symmetry of scalar field briefly. Then, we will introduce Lunin's result on the separation of variables for Maxwell equations on Kerr background.
Finally, we will discuss the hidden conformal symmetry of vector field based on that equation.

\subsection{Hidden conformal symmetry for scalar field}
Consider the Kerr metric for a stationary and axisymmetric black hole with mass \(M\) and angular momentum \(J=M a\),
\begin{align}
\label{gsam}
ds^2 = \dfrac{\Sigma}{\Delta}dr^2-\dfrac{\Delta}{\Sigma}(dt-a \sin^2\theta d\phi)^2+\Sigma d\theta^2+\dfrac{\sin^2\theta}{\Sigma}((r^2+a^2)d\phi-a dt)^2,
\end{align}
where \(\Delta\) and \(\Sigma\) are given by
\begin{align}
\Delta=r^2+a^2-2 M r, \ \ \
\Sigma=r^2+a^2\cos^2\theta.\notag
\end{align}
Consider the Klein-Gorden equation for a massless scalar field:
\begin{align}
\Box \Phi=0.
\end{align}
Since there exist two Killing vectors, \(\partial_t\) and \(\partial_\phi\),
the scalar field can be expanded in eigenmodes as
\begin{align}
\Phi(t,r,\theta,\phi)=e^{-i \omega t+i m \phi}\Phi(r,\theta).
\end{align}
By introducing the ansatz:
\begin{align}
\Phi(r,\theta)=R(r)S(\theta),
\end{align}
the Klein-Gorden equation can be separated into:
\begin{align}
\label{sawq}
[\frac{1}{\sin\theta}\partial_\theta(\sin\theta \partial_\theta)-\frac{m^2}{\sin^2\theta}+\omega^2a^2\cos^2\theta]S(\theta)=-K_lS(\theta),
\end{align}
and
\begin{align}
\label{srwq}
[\partial_r\Delta\partial_r+\frac{(2Mr_+\omega-am)^2}{(r-r_+)(r_+-r_-)}-\frac{(2Mr_-\omega-am)^2}{(r-r_-)(r_+-r_-)}+(r^2+2M(r+2M))\omega^2]R(r)=K_lR(r).
\end{align}
Then, under low frequency approximation:
\begin{align}
\omega M\ll 1,
\end{align}
the whole spacetime can be divided into two regions:
\begin{align}
\text{``Near\ Region''}\ \ \ r\ll \frac{1}{\omega},\\
\text{``Far \ Region''}\ \ \ r\gg M.
\end{align}
And there exist an overlap in the matching region:
\begin{align}
\text{``Matching\ Region''}\ \ \ M\ll r\ll \frac{1}{\omega}.
\end{align}

In the near region, under the condition \(r\ll \frac{1}{\omega}\), the angular equation \eqref{sawq} degenerates into a simpler form:
\begin{align}
[\frac{1}{\sin\theta}\partial_\theta(\sin\theta \partial_\theta)-\frac{m^2}{\sin^2\theta}]S(\theta)=-K_lS(\theta).
\end{align}
This equation is Legendre equation with separation constant \(K_l=l(l+1)\),
and the solutions $e^{im\phi}S(\theta)$ are spherical harmonics.

Then we focus on the radical equation \eqref{srwq} in the near region under low frequency approximation, which becomes
\begin{align}
\label{srwqlf}
[\partial_r\Delta\partial_r+\frac{(2Mr_+\omega-am)^2}{(r-r_+)(r_+-r_-)}-\frac{(2Mr_-\omega-am)^2}{(r-r_-)(r_+-r_-)}]R(r)=l(l+1)R(r).
\end{align}
The solutions to this equation are hypergeometric functions, and they transform in representations of \(SL(2,\mathbb{R})\).

Consider the coordinate transformation which maps the Boyer-Lindquist coordinates Eq. \eqref{gsam} to the so call `conformal coordinates',
\begin{align}
w^+=\sqrt{\frac{r-r_+}{r-r_-}}e^{2\pi T_R \phi},w^-=\sqrt{\frac{r-r_+}{r-r_-}}e^{2\pi T_L \phi-\frac{t}{2M}},y= \sqrt{\frac{r_+-r_-}{r-r_-}}e^{\pi(T_R +T_L) \phi-\frac{t}{4M}},
\end{align}
and
\begin{align}
T_R=\frac{r_+-r_-}{4\pi a},\ \ \ \ \ T_L=\frac{r_++r_-}{4\pi a}.
\end{align}
Then we can define six vector fields as:
\begin{align}\label{vfs}
H_1=i\partial_+, \ \ \ H_0=i(w^+\partial_++\frac{1}{2}y\partial_y), \ \ \ H_{-1}=i((w^+)^2\partial_++w^+y\partial_y-y^2\partial_-),\\
\bar{H}_1=i\partial_-, \ \ \ \bar{H}_0=i(w^-\partial_-+\frac{1}{2}y\partial_y), \ \ \ \bar{H}_{-1}=i((w^-)^2\partial_-+w^-y\partial_y-y^2\partial_-).
\end{align}
They obey the \(SL(2,\mathbb{R})\) Lie bracket algebra
\begin{align}
[H_m,H_n]=(m-n)H_{m+n},\\
[\bar{H}_m,\bar{H}_n]=(m-n)\bar{H}_{m+n}.
\end{align}
Then the corresponding \(SL(2,\mathbb{R})\) quadratic Casimir operator is:
\begin{align}
\label{casimirk}
\mathcal{H}^2=\bar{\mathcal{H}}^2=-H_0^2+\frac{1}{2}(H_1H_{-1}+H_{-1}H_1)
\end{align}
In (\(t,r,\theta,\phi\)) coordinates those vector fields can be expressed as:
\begin{align}
\label{Hoc}
&H_1=ie^{-2\pi T_R \phi}(\sqrt{\Delta}\partial_r+\frac{1}{2\pi T_R}\frac{r-M}{\sqrt{\Delta}}\partial_\phi+\frac{2T_L}{T_R}\frac{Mr-a^2}{\sqrt{\Delta}}\partial_t)\notag,\\
&H_0=\frac{i}{2\pi T_R}\partial_\phi+2iM\frac{T_L}{T_R}\partial_t,\notag\\
&H_{-1}=ie^{2\pi T_R \phi}(-\sqrt{\Delta}\partial_r+\frac{1}{2\pi T_R}\frac{r-M}{\sqrt{\Delta}}\partial_\phi+\frac{2T_L}{T_R}\frac{Mr-a^2}{\sqrt{\Delta}}\partial_t),
\end{align}
and
\begin{align}
\label{Hboc}
&\bar{H}_1=ie^{-2\pi T_L\phi+\frac{t}{2M}}(\sqrt{\Delta}\partial_r-\frac{a}{\sqrt{\Delta}}\partial_\phi-\frac{2Mr}{\sqrt{\Delta}})\notag,\\
&\bar{H}_0=-2iM\partial_t,\notag\\
&\bar{H}_{-1}=ie^{2\pi T_L\phi-\frac{t}{2M}}(-\sqrt{\Delta}\partial_r-\frac{a}{\sqrt{\Delta}}\partial_\phi-\frac{2Mr}{\sqrt{\Delta}}).
\end{align}
If we substituting Eq. \eqref{Hoc} and Eq. \eqref{Hboc} into Eq. \eqref{casimirk},  the Casimir operator will become
\begin{align}
\mathcal{H}^2=\partial_r\Delta\partial_r-\frac{(2Mr_+\partial_t+a\partial_\phi)^2}{(r-r_+)(r_+-r_-)}+\frac{(2Mr_-\partial_t+a\partial_\phi)^2}{(r-r_-)(r_+-r_-)}.
\end{align}
Then the radical equation in the near region Eq. \eqref{srwqlf} can be written as
\begin{align}
\bar{\mathcal{H}}^2\Phi=\mathcal{H}^2\Phi=l(l+1)\Phi.
\end{align}
So we can see that the equation of motion for massless scalar field can be written as the \(SL(2,\mathbb{R})\) Casimir.
And with the corresponding temperature \((T_R,T_L)\) and two central charges \((c_L,c_R)\)
which derived from the analysis of Kerr/CFT\cite{Guica:2008mu,Castro:2009jf} , we can reproduce the Bekenstein-Harking entropy using Cardy formula $S=\frac{\pi^2}{3}(c_LT_L+c_RT_R)$.
This means that there exist hidden conformal symmetry for a general Kerr black hole \cite{Castro:2010fd}.

\subsection{Separation of variables for Maxwell equations}

Consider the homogeneous (source-free) Maxwell equations for a massless vector field \(A^\mu\):
\begin{align}
\nabla_\mu F^{\mu \nu}=0,~~~\text{with}~~~F^{\mu \nu}=\nabla^\mu A^\nu-\nabla^\nu A^\mu.
\end{align}
According to \cite{Lunin:2017drx}, the Maxwell equations can be separated on the Kerr background under the assumptions:
\begin{align}
\label{ansatz}
l^\mu_\pm A_\mu=\pm \frac{r}{1\pm i \lambda r}l^\mu \partial_\mu \Phi, \  m^\mu_\pm A_\mu=\mp \frac{ia \cos\theta}{1\pm \lambda a \cos\theta}m^\mu \partial_\mu \Phi, \  \Phi=e^{-i\omega t+i m \phi}R(r)S(\theta),
\end{align}
where \(\lambda\) is an additional parameter, while \(l^\mu_\pm\) and \(m^\mu_\pm\) are related to the null tetrads \((l,n,m,\bar{m})\) of Kerr metric according to
\begin{align}
\label{tetrad}
&l^\mu_+=l^\mu,~l^\mu_-=-\frac{2\Sigma}{\Delta}n^\mu,~ m^\mu_+=\sqrt{2}\rho m^\mu,~ m^u_-=\sqrt{2}\bar{\rho}\bar{m}^\mu, \\
&\rho=r+ia \cos\theta,~~~\bar{\rho}=r-ia\cos\theta\notag.
\end{align}
We can easily find that the vectors \(l^\mu_\pm\) are not the functions of \(\theta\) and \(m^\mu_\pm\) are not the functions of \(r\):
\begin{align}
\label{lmpn}
l^\mu_\pm \partial_\mu=\partial_r\pm(\frac{r^2+a^2}{\Delta}\partial_t+\frac{a}{\Delta}\partial_\phi),~~~~m^\mu_\pm\partial_\mu=\partial_\theta\pm(ia\sin\theta \partial_t+\frac{i}{\sin\theta}\partial_\phi).
\end{align}
Taking Eq. \eqref{lmpn} into  Eq. \eqref{ansatz}, we can get the components of the vector field:
\begin{flalign}
\begin{split}
A^t=-\frac{\lambda R(r)S(\theta)}{\Sigma}e^{-i\omega t+im\phi}[&\frac{r^2(r^2+a^2)}{\Delta (1+\lambda^2r^2)}(a(m-a\omega)-\omega r^2+\frac{\Delta R'(r)}{\lambda r R(r)})\notag\\&+\frac{a^3\cos^2\theta}{1-a^2\lambda^2\cos^2\theta}(m-a\omega\sin^2\theta-\frac{\sin^2\theta S'(\theta)}{a\lambda \cos\theta S(\theta)})],\notag
\end{split}&
\end{flalign}
\begin{flalign}
\begin{split}
&A^r=\frac{irR(r)S(\theta)}{(1+\lambda^2r^2)\Sigma}e^{-i\omega t+im\phi}(a(m-a\omega)-\omega r^2-\frac{\lambda r \Delta R'(r)}{R(r)}),\notag\\
&A^\theta=\frac{ia\cos\theta R(r)S(\theta)}{(1-a^2\lambda^2\cos^2\theta)\Sigma}e^{-i\omega t+im\phi}(a\omega\sin^2\theta-m+\frac{a\lambda \cos\theta\sin^2\theta S'(\theta)}{S(\theta)})\notag,
\end{split}&
\end{flalign}
\begin{flalign}
\begin{split}
A^\phi=-\frac{a\lambda R(r)S(\theta)}{\Sigma}e^{-i\omega t+im\phi}[&\frac{r^2}{\Delta(1+\lambda^2r^2)}(-\omega r^2+a(m-a\omega)+\frac{\Delta R'(r)}{\lambda r R(r)})\notag\\&-\frac{a \cos^2\theta}{\sin^2\theta(1-a^2\lambda^2\cos^2\theta)}(a\omega\sin^2\theta-m+\frac{\sin^2\theta S'(\theta)}{a \lambda \cos\theta S(\theta)})].
\end{split}&
\end{flalign}
Substituting the above results into Maxwell equations, we can do the separation of variables as:
\begin{align}
\label{ROSV}
\frac{E_\theta}{\sin\theta}\frac{d}{d\theta}[\frac{\sin\theta}{E_\theta}\frac{dS(\theta)}{d\theta}]+(\frac{-2\Lambda}{E_\theta}+4a\omega+(a\omega \cos\theta)^2-\frac{m^2}{\sin^2\theta}-C)S(\theta)=0,\notag\\
E_r\frac{d}{dr}[\frac{\Delta}{E_r}\frac{dR(r)}{dr}]+(\frac{2\Lambda}{E_r}+r^2\omega^2-\frac{4am\omega\chi}{\Delta}+\frac{a^2m^2}{\Delta}+\frac{2Mr\omega^2\chi}{\Delta}+\frac{4Marm\omega}{\Delta}+C)R(r)=0.
\end{align}
For simplicity of the expressions, we have defined several functions
\begin{align}
&E_r=1+\lambda^2r^2,~~~E_\theta=1-(\lambda a \cos\theta)^2,~~~\chi=r^2+a^2,~~~\notag\\&\Lambda=a\lambda (m-a\omega+\frac{\omega}{a\lambda^2}),~~~C=-\Lambda+2am\omega+a^2\omega^2,
\end{align}
with the separation constant \(\lambda_1=-\frac{\omega}{\lambda}\).

This method to separate the Maxwell equation is different from the Teukolsky equation. And the Newman-Penrose scalars read:
\begin{align}
\Psi_0&=\frac{e^{-i\omega t+im\phi}\lambda_1^2}{\sqrt{2}}\frac{((r^2+a^2)\omega-am)R(r)+i\Delta R'(r)}{\Delta(i\lambda_1+r\omega)}\frac{(m-a\omega\sin^2\theta)S(\theta)+\sin^2\theta S'(\theta)}{\sin\theta(a\omega\cos\theta-\lambda_1)},\\
\Psi_2&=\frac{e^{-i\omega t+im\phi}\lambda_1^2}{2\sqrt{2}(ir+a\cos\theta)^2}\frac{((r^2+a^2)\omega-am)R(r)+i\Delta R'(r)}{分\lambda_1+ir\omega}\frac{(m-a\omega\sin^2\theta)S(\theta)-\sin^2\theta S'(\theta)}{\sin\theta(a\omega\cos\theta+\lambda_1)}
\end{align}

\subsection{Hidden conformal symmetry for vector field}

Now we focus on the near region under low frequency approximation, then the angular equation will be simplified into Legendre equation:
\begin{align}
[\frac{1}{\sin\theta}\partial_\theta(\sin\theta \partial_\theta)-\frac{m^2}{\sin^2\theta}+\lambda_1]S(\theta)=0.
\end{align}

The solutions to the above equation are spherical harmonics with separation constants \(\lambda_1=l(l+1)\). Note that this equation is different from the Teukolsky equation for vector perturbation, this equation does not contain spin weight. To our knowledge, this phenomenon may just because the target of this method is the vector field \(A^\mu\) not the field strength (\(\Psi_0,\Psi_2\)), and field strengths are functions of the first derivation of field, which is similar with the relation of spherical harmonics and spin weight spherical harmonics.

Then we are going to analyze the radial wave equation in the near region. It is necessary to note that \(\Delta\ll 1\) is also a small quantity in the near region, so the terms which contain \(\omega \) and \(\omega^2\) can not be neglected when they have the factor \(\Delta\) in denominator. This implies:
\begin{align}
\label{eor}
\frac{d}{dr}[\Delta\frac{dR(r)}{dr}]+(-\lambda_1-\frac{4am\omega\chi}{\Delta}+\frac{a^2m^2}{\Delta}+\frac{2Mr\omega^2\chi}{\Delta}+\frac{4Marm\omega}{\Delta})R(r)=0.
\end{align}
Notice that:
\begin{align}
&-\frac{4am\omega\chi}{\Delta}+\frac{4Marm\omega}{\Delta}=-\frac{4am\omega(\Delta+2Mr)}{\Delta}+\frac{4Marm\omega}{\Delta}=-4amw-\frac{4Marm\omega}{\Delta},\\
&\frac{2Mr\omega^2(r^2+a^2)}{\Delta}=\frac{2Mr\omega^2(\Delta+2Mr)}{\Delta}=2Mr\omega^2+\frac{4M^2r^2w^2}{\Delta},\\
&\frac{4M^2r^2\omega^2}{\Delta}=\frac{4M^2\omega^2\Delta+8M^3rw^2-4M^2a^2\omega^2}{\Delta}=4M^2w^2+\frac{8M^3rw^2-4M^2a^2\omega^2}{\Delta}.
\end{align}
Taking those three relations into account, then in the near region:
\begin{align}
&\frac{a^2m^2}{\Delta}-\frac{4am\omega\chi}{\Delta}+\frac{2Mr\omega^2\chi}{\Delta}+\frac{4Marm\omega}{\Delta}=\frac{(2Mr_+\omega-am)^2}{(r-r_+)(r_+-r_-)}-\frac{(2Mr_-\omega-am)^2}{(r-r_-)(r_+-r_-)},\notag\\
&r_+=M+\sqrt{M^2-a^2},\ \ \ \ \ r_-=M-\sqrt{M^2-a^2}.
\end{align}
Then the radial equation in the near region Eq. \eqref{eor} becomes:
\begin{align}
\label{eornr}
\frac{d}{dr}[\Delta\frac{dR(r)}{dr}]+(\frac{(2Mr_+\omega-am)^2}{(r-r_+)(r_+-r_-)}-\frac{(2Mr_-\omega-am)^2}{(r-r_-)(r_+-r_-)}-l(l+1))R(r)=0.
\end{align}
This equation is exactly the same with Eq. \eqref{srwqlf}. So we can conclude that there also exist hidden conformal symmetry for Maxwell field on Kerr background, and the temperatures \((T_L,T_R)\) are exactly equal to the temperatures derived from the scalar situation \cite{Castro:2010fd}. Then, with Cardy formula, we can get Bekenstein-Hawking entropy exactly the same as the scalar case.

\section{Vector field on Kerr-NUT-(A)dS background}
In this section, we will separate the Maxwell equation in Kerr-NUT-(A)dS geometry, and then we impose the low frequency approximation in the near horizon region, we find that there still exist hidden conformal symmetry of separation equation of Maxwell equation in this background. The details of the corresponding analyses are similar to Kerr, so we just present necessary steps and main results.

\subsection{Separation of variables for vector field}
Kerr-NUT-(A)dS metrics are one kind of generalizations of Kerr solutions with  a NUT parameter \cite{Chen:2006xh} in the presence of cosmological constant. In the Boyer-Lindquist type coordinates, this metric reads:
\begin{align}
ds^2=-\frac{\Delta f^{KNA}_r}{\Sigma}(dt-\frac{a\sin^2\theta}{\Xi}d\phi)^2+\frac{\Sigma}{\Delta f^{KNA}_r}dr^2+\frac{\Sigma a^2\sin^2\theta}{f^{KNA}_\theta}d\theta^2+\frac{f^{KNA}_\theta}{\Sigma}(dt-\frac{r^2+a^2}{a \Xi}d\phi)^2.
\end{align}
where
\begin{align}
&f^{KNA}_r=1-\frac{Lr^2(r^2+a^2)}{\Delta},\\
&f^{KNA}_\theta=a^2\sin^2\theta(1+La^2\cos^2\theta)+2na\cos\theta.
\end{align}
\(L\) is related to the cosmological constant, and \(n\) is the NUT parameter, this metric degenerate into Kerr-(A)dS \cite{Carter:1968ks} with setting \(n=0\).

Following the same ansatz in Eq.\eqref{ansatz} with different null tetrads, and substitute those null tetrads of Kerr-NUT-(A)dS into the relation \eqref{tetrad}, we can get
\begin{align}
&l^\mu_\pm\partial_\mu=\sqrt{f^{KNA}_r}\partial_r\pm\frac{1}{\sqrt{f^{KNA}_r} \Delta}((r^2+a^2)\partial_t+a\Xi\partial_\phi ),\notag\\
&m^\mu_\pm\partial_\mu=\frac{\sqrt{f^{KNA}_\theta}}{a}\partial_\theta\pm\frac{ia}{\sqrt{f^{KNA}_\theta}}(a\sin\theta\partial_t+\frac{\Xi}{\sin\theta}\partial_\phi).
\end{align}
The components of vector-potential can be gotten by substituting the above four vectors into the ansatz. \eqref{ansatz}. And taking those vector-potential components into Maxwell equation, we will get two separated equations:
\begin{align}
\label{eoswnads}
\frac{E_\theta}{\sin\theta}\frac{d}{d\theta}[\frac{f^{KNA}_\theta}{a^2\sin\theta E_\theta}\partial_\theta S(\theta)]+[-\frac{(a\sin\theta)^4}{f^{KNA}_\theta}(-\omega+\frac{m\Xi}{a\sin^2\theta})^2-\frac{2\Lambda_1}{E_\theta}+\Lambda_1]S(\theta)=0,
\end{align}
\begin{align}
\label{eorwnads}
E_r\frac{d}{dr}[\frac{f^{KNA}_r\Delta}{E_r}\partial_r R(r)]+[\frac{1}{f^{KNA}_r\Delta}(-\omega(r^2+a^2)+am\Xi)^2+\frac{2\Lambda_1}{E_r}-\Lambda_1]R(r)=0.
\end{align}
And \(E_\theta, E_r\) and \(\Lambda_1\) have been defined clearly above \footnote{When we prepare this paper, Frolov et al \cite{Frolov:2018pys} do separate Maxwell equation in this background with different coordinates, these two sets of equations  are the same under coordinate transformation.}.

\subsection{Low frequency limit at Near-Horizon region}
Similar with the discussion we did in Kerr background, we will focus on the low frequency limit with \(\omega M\ll 1\). Then the angular equation degenerates into
\begin{align}
\label{eosnads}
\frac{E_\theta}{\sin\theta}\frac{d}{d\theta}[\frac{f^{KNA}_\theta}{a^2\sin\theta E_\theta}\partial_\theta S(\theta)]+[-\frac{a^2 m^2\Xi^2}{f^{KNA}_\theta}+\lambda_1]S(\theta)=0,
\end{align}
Although it is hard to solve this equation analytically, we can get an vital result that \(\lambda_1\) doesn't depend on \(\omega\), which is enough to discuss the hidden symmetry of radical equation.

In virtue of the function that locate the horizon is no longer quadratic, we can't discuss the hidden conformal symmetry as we did in the Kerr spacetime. Similar with the analysis in \cite{Chen:2010bh}, we focus on the near-horizon region, where \(r-r_+\) is a small quantity, so we can series \(f_r^{KNA} \Delta\)  to the quadratic order,
\begin{align}
f^{KNA}_r\Delta=(1-L r^2)(r^2+a^2)-2Mr\simeq c_3(r-r_+)(r-r_*),
\end{align}
where \(r_+\) always stands for outer horizon, and
\begin{align}
&c_3=1-La^2-6L r_+^2,\notag\\
&r_*=r_+-\frac{1}{c_3 r_+}(r_+^2-a^2-L(3r_+^4+a^2r_+^2)).\notag
\end{align}

Taking those approximations into consideration, the radical equation can be simplified in the near-horizon region:
\begin{flalign}
\label{eornradskna}
\partial_r(r-r_+)(r-r_*)&\partial_rR(r)\notag\\+&(\frac{(ma\Xi-\omega(r_+^2+a^2))^2}{c_3^2(r-r_+)(r_+-r_*)}-\frac{(ma\Xi-\omega(r_*^2+a^2))^2}{c_3^2(r-r_*)(r_+-r_*)}-\frac{\lambda_1}{c_3})R(r)=0,
\end{flalign}
and we will see in the next subsection that this wave equation can be reproduce by \(SL(2,\mathbb{R})\) Casimir.

\subsection{Hidden conformal symmetry for vector field}
Similar with the logic that we discuss in the Kerr geometry, we will introduce the `conformal coordinate':
\begin{align}
w^+=\sqrt{\frac{r-r_+}{r-r_*}}e^{2\pi T_R \phi},w^-=\sqrt{\frac{r-r_+}{r-r_*}}e^{2\pi T_L \phi-\frac{t}{r_++r_*}},y= \sqrt{\frac{r_+-r_*}{r-r_*}}e^{\pi(T_R +T_L) \phi-\frac{t}{2(r_++r_*)}},
\end{align}
and
\begin{align}
T_R=\frac{c_3(r_+-r_*)}{4\pi a\Xi},\ \ \ \ \ T_L=\frac{c_3(r_+^2+r_*^2+2a^2)}{4\pi a\Xi(r_++r_*)}.
\end{align}
Defining six vector fields which are exactly the same form with Eq. \eqref{vfs}, and they possess \(SL(2,R)_R\times SL(2,R)_L\) respectively.

The related Casimir operator can be expressed in (\(t,r,\theta,\phi\)) coordinates as:
\begin{align}
\mathcal{H}^2=\partial_r(r-r_+)(r-r_*)&\partial_r-\frac{(a\Xi\partial_\phi+(r_+^2+a^2)\partial_t)^2}{c_3^2(r-r_+)(r_+-r_*)}-\frac{(a\Xi\partial_\phi+(r_*^2+a^2)\partial_t)^2}{c_3^2(r-r_*)(r_+-r_*)},
\end{align}
then, Eq. \eqref{eornradskna} can be written:
\begin{align}
\mathcal{H}^2\Phi=\frac{\lambda_1}{c_3}\Phi.
\end{align}

And from the general argument in \cite{Mei:2010wm}, for the case of Kerr-NUT-(A)dS geometry, the central charge should be:
\begin{align}
c_L=c_R=\frac{6a(r_++r_*)}{c_3}.
\end{align}
The Cardy formula gives the microscopic entropy:
\begin{align}
S_m=\frac{\pi^2}{3}(c_LT_L+c_RT_R)=\pi\frac{a^2+r_+^2}{1+L a^2}=S_{BH}.
\end{align}
Thus, we can conclude that there also exist the hidden conformal symmetry of separation equation of Maxwell field in Kerr-NUT-(A)dS spacetime, and the corresponding temperatures together with the central charges reproduce the Bekenstein-Hawking entropy.

\section{Conclusion}
In this paper, we have extended the result  on separation of variables for Maxwell equation on Kerr black hole to the more general Kerr-NUT-(A)dS black hole. Then under the low frequency approximation, and restrict to the near or near-horizon region,
the separated equations will be simplified to the same form as scalar field.
So it's very clear to see that there exist hidden conformal symmetry for vector field on those backgrounds.

We have revealed the hidden conformal symmetry for vector field on different black holes: Kerr and Kerr-NUT-(A)dS.
We also checked the black holes with electric charge, and the results are the same.
Since the logics are very similar to the above discussions, we will not list all the results here.
We speculate that for other black hole backgrounds,
the separation of variables for Maxwell equations can also be achieved following the same method.
And it's reasonable to conjecture that under proper coordinates and suitable approximations,
the vector equation will be simplified to the same form.

Up to now, the hidden conformal symmetry has been extended to vector field,
but whether this property will hold for tensor field is still an open question.
To answer this question, a separable equation for tensor field should be found first.
Following the work of vector field, it may be achieved by a careful study of solutions to Teukolsky equations.
But for tensor field, it's difficult to find out the suitable ansatz for components of tensor field.

In the future work, we will focus on the tensor field.
If we can find hidden conformal symmetry for tensor field, it will be a stronger evidence of holographic principle.
Further more, it will be helpful when we are trying to solve the gravitational wave equations for systems such as extreme mass ratio inspirals (EMRIs).

\acknowledgments

This work has been supported by the Natural Science Foundation of China (Grant Nos. 91636111, 11475064, 11690022).


\end{document}